# Fast-Converging Deep Reinforcement Learning for Optimal Dispatch of Large-Scale Power Systems under Transient Security Constraints

Tannan Xiao, *Member*, *IEEE*, Ying Chen, *Senior Member*, *IEEE*, Han Diao, Shaowei Huang, *Member*, *IEEE*, and Chen Shen, *Senior Member*, *IEEE*

***Abstract*—Power system optimal dispatch with transient security constraints is commonly represented as Transient Security-Constrained Optimal Power Flow (TSC-OPF). Deep Reinforcement Learning (DRL)-based TSC-OPF trains efficient decision-making agents that are adaptable to various scenarios and provide solution results quickly. However, due to the high dimensionality of the state space and action spaces, as well as the non-smoothness of dynamic constraints, existing DRL-based TSC-OPF solution methods face a significant challenge of the sparse reward problem. To address this issue, a fast-converging DRL method for TSC-OPF is proposed in this paper. The Markov Decision Process (MDP) modeling of TSC-OPF is improved by reducing the observation space and smoothing the reward design, thus facilitating agent training. An improved Deep Deterministic Policy Gradient algorithm with Curriculum learning, Parallel exploration, and Ensemble decision-making (DDPG-CPEn) is introduced to drastically enhance the efficiency of agent training and the accuracy of decision-making. The effectiveness, efficiency, and accuracy of the proposed method are demonstrated through experiments in the IEEE 39-bus system and a practical 710-bus regional power grid. The source code of the proposed method is made public on GitHub.**

## I. Introduction

### A. Motivations

Power System Optimal Dispatch with Transient Security Constraints (PSOD-TSC) ensures the safe operation of power grids under certain anticipated contingencies by adjusting power flow distributions. However, with the continuous development of power systems, the uncertainty of operating states has significantly increased, posing serious challenges to PSOD-TSC.

PSOD-TSC is commonly modeled as the Transient Security-Constrained Optimal Power Flow (TSC-OPF) problem [1]. Since the introduction of this problem, researchers have proposed various methods to solve it [1]-[2]. Among them, Deep Reinforcement Learning (DRL)-based methods construct an agent using neural networks and train the agent by interacting with a simulation-based environment. After training, the decision-making by the agent is extremely fast, making it a highly promising solution for PSOD-TSC [3].

However, the application of DRL methods in large-scale power systems is significantly restricted because the methods converge slowly and often fail to obtain a well-trained agent. The slow convergence of agent training is due to the low efficiency of exploration caused by the *sparse reward problem* [4]. The sparse reward problem refers to the difficulty of obtaining positive rewards when the agent explores the action space. This problem arises due to two main factors. Firstly, both the state space and the action space are high-dimensional, which necessitates tens of thousands of samples to adequately cover these spaces. Secondly, the operation of a power grid is subject to physical laws, resulting in a limited feasible domain within the action space. Consequently, during exploration, the agent often generates infeasible actions that lead to non-convergent power flows, violations of static security constraints, or violations of dynamic security constraints. Thus, a significant increase in the number of samples is required for agents to learn good actions with positive rewards.

The sparse reward problem can cause the agent to repeatedly explore infeasible regions, thereby greatly impacting the convergence of agent training and potentially preventing it from learning effective strategies. Consequently, in literature, DRL-based TSC-OPF solution methods have been applied to systems with up to 140 nodes, employing a multi-agent algorithm [5].

Therefore, in this paper, a fast-converging DRL method is proposed to mitigate this sparse reward problem. The proposed method is verified in the IEEE 39-bus system and a practical 710-bus regional power grid. Related works are as follows.

### B. Related Works

TSC-OPF is a challenging nonlinear optimization problem, making it difficult to solve and verify the optimality of the solution. There are mainly four types of solution methods, namely, dynamic optimization-based methods [6], [7], [8], simplification-based methods [9], [10], meta-heuristics methods [11], [12], and data-driven methods. Generally, dynamic optimization-based methods offer high solution

This work was supported in part by the National Natural Science Foundation of China (NSFC) under Grants 52107104. (*Corresponding Author: Ying Chen*).

Tannan Xiao is with the State Key Laboratory of Power System Operation and Control, Tsinghua University, Beijing, 10084, China (Email: xiaotn@tsinghua.edu.cn).

Ying Chen, Han Diao, Shaowei Huang, and Chen Shen are with the Department of Electrical Engineering, Tsinghua University, Beijing, 100084, China (Email: chen_ying@tsinghua.edu.cn, dh22@mails.tsinghua.edu.cn, huangsw@tsinghua.edu.cn, shenchen@mail.tsinghua.edu.cn).

accuracy, but they are challenging to implement and may encounter convergence issues, especially in large-scale power systems. Simplification-based methods, on the other hand, have lower computational complexity, but their solutions tend to be conservative due to the lack of accurate and reliable model simplification. Meta-heuristics methods are highly adaptable and easy to implement. With enough population and iterations, these methods can often obtain a sufficiently accurate solution. However, they come with a high computational burden that significantly increases with the scale of power grids. This paper focus on solving TSC-OPF for large-scale power systems. Therefore, a Particle Swarm Optimization (PSO)-based solution method is utilized as the benchmark in the case study.

A common issue with dynamic optimization-based methods, simplification-based methods, and meta-heuristics methods is that the entire decision-making process needs to be performed for every different scenario. Therefore, a large amount of historical data is generated but not effectively utilized.

Naturally, with the development of artificial intelligence, data-driven methods such as deep learning and DRL, which can extract hidden patterns from massive historical data, have also been applied to TSC-OPF.

Deep learning-based methods use deep neural networks as surrogate models for some computationally expensive or challenging tasks in TSC-OPF solutions, such as generating stability criteria [13], calculating starting points of TSC-OPF [14], predicting effective security constraints [15], generating unit commitment solutions [16], generating components' local control strategies [17], generating current stage control strategies in multi-stage optimization problems [18], [19], etc. The prerequisite for the application of these methods is the existence of a large amount of historical data available.

DRL-based TSC-OPF solution methods train agents for TSC-OPF solution through interaction with simulation-based environments. The advantages of DRL methods are twofold. On the one hand, DRL methods search in the function space to find a policy function that can provide optimal TSC-OPF solutions for different scenarios with extreme efficiency. On the other hand, different from deep learning methods, the training of agents does not require labeled data samples prepared in advance. It only needs a corresponding environment for RL.

The limited exploration efficiency caused by the sparse reward problem is a major drawback of DRL-based methods. When both the static and dynamic security constraints are considered, agents often struggle to find actions that yield a positive reward within the feasible region. Consequently, in the existing literature, DRL methods are primarily employed for OPF solutions [20], [21], [22], [23], [24], rather than TSC-OPF solutions [25], [5]. In [25], the authors propose a neural barrier function to restrict the range of actions determined by the agent, which essentially serves as a mitigation strategy for the sparse reward problem. They successfully train an agent for the IEEE 118-bus system. In [5], a distributed Deep Deterministic Policy Gradient (DDPG)-based approach is used to train a preventive control agent for TSC-OPF in an asynchronous manner. A well-trained agent for the NPCC 140-bus system is obtained. Although parallel computing-based asynchronous training and multi-agent algorithms facilitate efficient exploration of the action space, they come with challenges such as difficulties in designing asynchronous training schemes and coordinating multi-agent optimization.

### C. Contributions

In this paper, a fast-converging DRL method for large-scale PSOD-TSC is proposed. The contributions are as follows.

1) An improved Markov Decision Process (MDP) modeling of TSC-OPF is developed to facilitate agent training. A general MDP modeling scheme for TSC-OPF is introduced to reduce the observation space. A four-stage smooth reward design based on the simulation time duration of instability, hereafter referred to as the *instability duration*, is adopted to quantify the level of instability.

2) An enhanced DDPG algorithm based on Curriculum learning, Parallel exploration, and Ensemble decision-making (DDPG-CPEn) is proposed to ensure stable and rapid convergence of agent training. The incorporation of a curriculum learning stage enhances the agent's ability to generate actions that result in convergent power flow, thus speeding up the initial training of the agent. A parallel exploration technique is introduced to quickly expand the replay buffer and improve the efficiency of exploring the action space. Final solutions are obtained through ensemble decision-making of five well-trained agents, balancing accuracy and efficiency.

3) A fast-converging DRL method for PSOD-TSC is implemented using Python and is publicly available on GitHub [26]. The method is validated in the IEEE 39-bus system and a practical 710-bus regional power grid, both utilizing detailed dynamic models. Notably, the training process exhibits stable and rapid convergence. In 10,000 new scenarios, the proposed method achieves success rates of 99.80 percent and 100.00 percent, respectively. The proposed method achieves a desirable trade-off between accuracy and efficiency by attaining average rewards of 96.12 percent and 97.32 percent and average time costs of 0.02 percent and 0.02 percent when compared to those of the Particle Swarm Optimization (PSO) algorithm, respectively.

### D. Paper Organization

The remainder of the paper is as follows. Section II introduces the TSC-OPF modeling of PSOD. The improved MDP modeling is introduced in Section III. Section IV demonstrates the detailed procedures of the DDPG-CPEn algorithm. Case studies are carried out in Section V. Finally, conclusions are drawn in Section VI.

## II. Problem Formulation of PSOD-TSC

### A. TSC-OPF Modeing

As mentioned before, PSOD-TSC is mathematically modeled as the TSC-OPF problem, which involves solving a nonlinear programming problem with constraints of DAEs that capture the dynamics of power systems. In this paper, the mathematical model of TSC-OPF is shown in (1):

$$
\begin{aligned}
&\min \quad C\left(\boldsymbol{x}_0, \boldsymbol{y}_0, \boldsymbol{u}\right) \\
&\text{s.t.} \quad \begin{cases}
\boldsymbol{g}\left(\boldsymbol{x}_0, \boldsymbol{y}_0, \boldsymbol{u}\right) = \boldsymbol{0} \\
\boldsymbol{h}\left(\boldsymbol{x}_0, \boldsymbol{y}_0, \boldsymbol{u}\right) \geq \boldsymbol{0} \\
\boldsymbol{x}(0) = \boldsymbol{x}_0, \boldsymbol{y}(0) = \boldsymbol{y}_0 \\
\boldsymbol{\Psi}\left[\dot{\boldsymbol{x}}(t), \boldsymbol{x}(t), \boldsymbol{y}(t), \boldsymbol{u}; \gamma\right] = \boldsymbol{0}, \forall t \in \left[0, T_E\right], \forall \gamma \in \boldsymbol{\Gamma} \\
\boldsymbol{\Phi}\left[\boldsymbol{x}(t), \boldsymbol{y}(t), \boldsymbol{u}; \gamma\right] = \boldsymbol{0}, \forall t \in \left[0, T_E\right], \forall \gamma \in \boldsymbol{\Gamma} \\
\boldsymbol{\varphi}\left[\boldsymbol{x}(t), \boldsymbol{y}(t); \gamma\right] \geq \boldsymbol{0}, \forall t \in \left[0, T_E\right], \forall \gamma \in \boldsymbol{\Gamma}
\end{cases}
\end{aligned} \tag{1}
$$

where $C$ denotes the objective function, $\boldsymbol{x}$ represents the state vector, whose time derivatives equal to $\dot{\boldsymbol{x}}$, $\boldsymbol{y}$ denotes the operation vector, $\boldsymbol{u}$ represents the control vector, $\boldsymbol{g}$ denotes the static equality constraints of power flow equations, $\boldsymbol{h}$ represents the static security constraints including nodal voltage limits, active and reactive generation limits, and transmission power limits, etc., the subscript $0$ represents the steady state value of the vectors, $T_E$ denotes the total simulation time, the $(t)$ following the vectors represent their values at the specific time instant $t$, $\gamma$ represents an anticipated contingency and all these contingencies form an anticipated contingency set $\boldsymbol{\Gamma}$, $\boldsymbol{\Psi}$ denotes the differential equations of DAEs, $\boldsymbol{\Phi}$ denotes the algebraic equations of DAEs, and $\boldsymbol{\varphi}$ represents the dynamic security constraints. This paper focuses on transient security. Therefore, the adopted transient security constraint is that the maximum rotor angle difference should not exceed 180 degrees during the entire simulation, as shown in (2):

$$
\pi - \Delta\delta_{\max}(t;\gamma) \geq \boldsymbol{0}, \forall t \in \left[0, T_E\right], \forall \gamma \in \boldsymbol{\Gamma} \tag{2}
$$

### B. Details of TSC-OPF Modeling

Constraint transcription methods are commonly employed to tackle TSC-OPF shown in (1). These methods convert the dynamic constraints throughout the simulation into the constraints solely at the end of the simulation, thus decoupling the optimization solution from the DAEs solution, as in (3):

$$
\begin{aligned}
&\min \quad C\left(\boldsymbol{x}_0, \boldsymbol{y}_0, \boldsymbol{u}\right) \\
&\text{s.t.} \quad \begin{cases}
\boldsymbol{g}\left(\boldsymbol{x}_0, \boldsymbol{y}_0, \boldsymbol{u}; \boldsymbol{\mathcal{E}}\right) = \boldsymbol{0} \\
\boldsymbol{h}\left(\boldsymbol{x}_0, \boldsymbol{y}_0, \boldsymbol{u}; \boldsymbol{\mathcal{E}}\right) \geq \boldsymbol{0} \\
\left[\boldsymbol{x}(T_E), \boldsymbol{y}(T_E); \boldsymbol{\mathcal{E}}, \gamma\right] = \boldsymbol{\mathcal{T}}\left(\boldsymbol{x}_0, \boldsymbol{y}_0, \boldsymbol{u}; \boldsymbol{\mathcal{E}}, \gamma\right), \forall \gamma \in \boldsymbol{\Gamma} \\
\boldsymbol{\varphi}'\left[\boldsymbol{x}(T_E), \boldsymbol{y}(T_E); \boldsymbol{\mathcal{E}}, \gamma\right] \geq \boldsymbol{0}, \forall \gamma \in \boldsymbol{\Gamma}
\end{cases}
\end{aligned} \tag{3}
$$

where $\boldsymbol{\varphi}'$ represents the dynamic security constraints after transcription and $\boldsymbol{\mathcal{T}}\left(\boldsymbol{x}_0, \boldsymbol{y}_0, \boldsymbol{u}; \gamma\right)$ denotes a function of $\boldsymbol{x}_0$, $\boldsymbol{y}_0$ and $\boldsymbol{u}$ given $\gamma$. The DAEs, which are solved using an external power system time-domain simulator, are implicitly contained in $\boldsymbol{\mathcal{T}}\left(\boldsymbol{x}_0, \boldsymbol{y}_0, \boldsymbol{u}; \gamma\right)$. As can be seen, the interaction between the optimizer and the external simulator is similar to the interaction between the agent and the RL environment. Therefore, TSC-OPF is modeled using the constraint transcription method in this paper. The details of (3) are as follows.

#### 1) Control vector $\boldsymbol{u}$

In this paper, the control vector $\boldsymbol{u}$ includes $\boldsymbol{V}_C$, the nodal voltage of all generator buses, and $\boldsymbol{P}_C$, the active generation of all the generators besides slack machines.

#### 2) Objective function $C$

This paper considers the objective of generation cost minimization, as shown in (4):

$$
\min \quad C = \text{sum}(\boldsymbol{C}_0) + \boldsymbol{C}_1^T \boldsymbol{P}_G + \boldsymbol{C}_2^T \left(\boldsymbol{P}_G \odot \boldsymbol{P}_G\right) \tag{4}
$$

where $\boldsymbol{C}_0$, $\boldsymbol{C}_1$, and $\boldsymbol{C}_2$ denote the coefficient vectors of generation cost, $\text{sum}(\boldsymbol{C}_0)$ calculates the summation of all the elements in $\boldsymbol{C}_0$, $\boldsymbol{P}_G$ denotes the active generation vector of generators, $\odot$ represents the Hadamard product, and $\boldsymbol{P}_G \odot \boldsymbol{P}_G$ denotes the element-wise square vector of $\boldsymbol{P}_G$. In this paper, elements in $\boldsymbol{C}_0$, $\boldsymbol{C}_1$, and $\boldsymbol{C}_2$ are set to 0.2, 30, and 100, respectively.

#### 3) Power flow constraints $\boldsymbol{g}$

$$
\begin{cases}
P_{Gi} - P_{Di} - V_i \sum_{j=1}^{N} V_j \left(G_{ij} \cos\theta_{ij} + B_{ij} \sin\theta_{ij}\right) = 0 \\
Q_{Gi} - Q_{Di} - V_i \sum_{j=1}^{N} V_j \left(G_{ij} \sin\theta_{ij} - B_{ij} \cos\theta_{ij}\right) = 0
\end{cases}, 0 \leq i, j < N_B \tag{5}
$$

where $P_{Gi}$ and $Q_{Gi}$ represent the active and reactive power injections at bus $i$ respectively, $P_{Di}$ and $Q_{Di}$ denote the active and reactive loads at bus $i$ respectively, $V_i$ and $V_i$ represent the nodal voltage amplitudes of buses $i$ and $j$ respectively, $G_{ii}$ and $B_{ii}$ respectively denote the conductance and susceptance of the branch between buses $i$ and $j$, $\theta_{ii}$ represents the voltage phase difference between buses $i$ and $j$, and $N_B$ is the total number of buses in the system.

#### 4) Static security constraints $\boldsymbol{h}$

$$
\begin{cases}
\underline{\boldsymbol{V}} \leq \boldsymbol{V} \leq \overline{\boldsymbol{V}} \\
\underline{\boldsymbol{P}}_G \leq \boldsymbol{P}_G \leq \overline{\boldsymbol{P}}_G \\
\underline{\boldsymbol{Q}}_G \leq \boldsymbol{Q}_G \leq \overline{\boldsymbol{Q}}_G \\
\underline{\boldsymbol{P}}_L \leq \boldsymbol{P}_L \leq \overline{\boldsymbol{P}}_L
\end{cases} \tag{6}
$$

where $\boldsymbol{V}$ denotes the nodal voltage vector, $\overline{\boldsymbol{P}}_G$ and $\underline{\boldsymbol{P}}_G$ represent the upper and lower limits of $\boldsymbol{P}_G$, $\overline{\boldsymbol{Q}}_G$ and $\underline{\boldsymbol{Q}}_G$ denote the upper and lower limits of $\boldsymbol{Q}_G$, and $\boldsymbol{P}_L$, $\overline{\boldsymbol{P}}_L$, and $\underline{\boldsymbol{P}}_L$ represent the active transmission power vector and its upper and lower limits, respectively.

#### 5) Transformed transient security constraints $\boldsymbol{\varphi}'$

$$
\pi - \Delta\delta_{\max}(T_E;\gamma) \geq 0, \forall \gamma \in \boldsymbol{\Gamma} \tag{7}
$$

where $\Delta\delta_{\max}(T_E;\gamma)$ represents the maximum rotor angle difference at the end of the simulation of the power system under contingency $\gamma$.

## III. MDP Modeling Improvements for TSC-OPF

### A. Basics of RL and MDP

The framework of RL is demonstrated in Fig. 1. State $\boldsymbol{s}$, action $\boldsymbol{a}$, and reward $r$ are the three basic factors of RL. A typical RL process involves the agent deciding on an action based on the state, the environment transitioning to a new state, the agent receiving a reward for the action, and then the agent updating its policy based on the received states and reward.

MDP is a standard mathematical modeling method for RL. MDP modeling requires the state transition process to have Markov property, which means that the current state contains all the information that affects the decision-making.

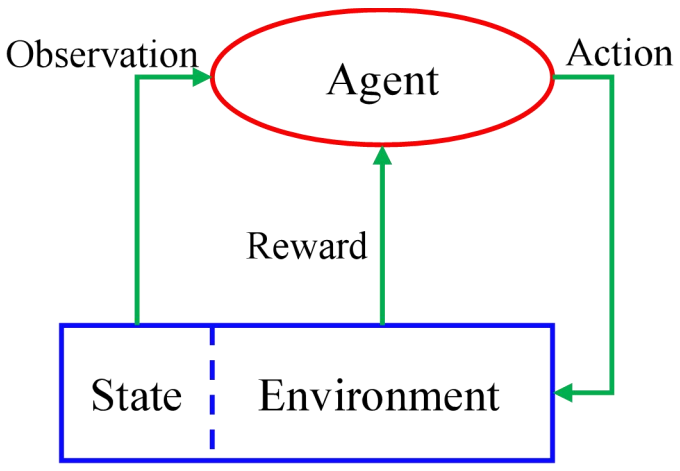

Fig. 1 Basic framework of RL

Other factors of RL include the observation $\boldsymbol{o}$, the policy $\pi$, the value function $v$, the reward discount factor $\gamma$, etc.

One of the key points in using RL to solve the TSC-OPF problem is to build an MDP model in a way that the objective of the MDP model is consistent with the original problem, the decision process has Markov property, and the hidden patterns in the environment are easy to learn.

### B. Observation Space Reduction

The observation $\boldsymbol{o}$ is closely related to the state $\boldsymbol{s}$. In general, $\boldsymbol{s}$ is a private representation of the environment that cannot be directly accessed, whereas the observation $\boldsymbol{o}$ is the information that the agent can actually obtain. The agent needs to obtain information about the environment through $\boldsymbol{o}$.

Reducing the dimensionality of the observation space is often advantageous for agent training. The control vector $\boldsymbol{u}$ of TSC-OPF determines the action space, i.e., $\boldsymbol{a} = \boldsymbol{u} \in [\boldsymbol{V}_C^T, \boldsymbol{P}_C^T]^T$. By decreasing the scale of the observation, the size of neural networks is directly reduced, thereby alleviating the challenges associated with model training and decreasing the required sample volume.

Therefore, a three-step MDP modeling scheme for DRL-based TSC-OPF solution methods is proposed to reduce the observation space. The principle of this scheme is to effectively reduce the observation space while maintaining the Markov property. The complete state of the environment, which strictly adheres to the Markov property, is gradually reduced to form the observation based on the engineering requirements and the characteristics of the original optimization problem. The specific steps involved in this scheme are outlined as follows.

Firstly, design a complete state that contains all the information of the environment, which gives the state transition strict Markov property. For TSC-OPF, the complete state includes power grid topology, power flow distribution, and parameters of dynamic components.

Secondly, the complete state is reduced to obtain a custom state by making MDP assumptions that align with engineering requirements. In this paper, we assume that the grid topology and dynamic parameters are known and remain constant. As a result, the custom state includes specific information that determines the distribution of power flow, i.e., $\boldsymbol{s} \in [\boldsymbol{V}_C^T, \boldsymbol{P}_C^T, \boldsymbol{P}_D^T, \boldsymbol{Q}_D^T]^T$. By defining this custom state, the actual sampling range in the environment is determined.

Finally, the custom state is further reduced based on the characteristics of the original optimization formulation, i.e., the TSC-OPF model shown in (1), to obtain the observation. The minimum generation cost is only determined by the load level. Therefore, the observation is further reduced to $\boldsymbol{o} \in [\boldsymbol{P}_D^T, \boldsymbol{Q}_D^T]^T$.

### C. Reward Design Smoothing

The design of rewards plays a crucial role in RL as it facilitates agents in learning hidden patterns within the environment. By defining appropriate rewards, the original optimization objective, along with its associated constraints, can be transformed into an equivalent problem of value function $v(\boldsymbol{o})$ maximization. In the case of TSC-OPF, the dynamic security constraints after transcription shown in (7) are non-smooth, which necessitates careful consideration during the reward design process.

In this paper, a four-stage reward shown in (8) is designed based on the instability duration that meets with (9):

$$r = \begin{cases} -1000, \text{non-convergent power flow} \\ \max\left(-500 - \boldsymbol{\lambda}_{dyn}^T \Delta \boldsymbol{T}_S, -999\right), \text{dynamic constraints violations} \\ \max\left(-\boldsymbol{\lambda}_{st}^T [\breve{\boldsymbol{V}}^T, \breve{\boldsymbol{P}}_G^T, \breve{\boldsymbol{Q}}_G^T, \breve{\boldsymbol{P}}_L^T]^T, -499\right), \text{static constraints violations} \\ \lambda_{opt} \times \left(1 - \dfrac{C}{\max(C)}\right), \text{no constraint violation} \end{cases} \tag{8}$$

$$\left(\Delta T_S; \gamma\right) = \left(T_E - T_S; \gamma\right) = 0, \forall \gamma \in \boldsymbol{\Gamma} \tag{9}$$

where $T_S$ is the time instant when the power system loses stability, i.e., the simulation time during which the system maintains stability following (2), $\Delta \boldsymbol{T}_S$ represents the instability duration vector of the anticipated contingency set $\boldsymbol{\Gamma}$, $\boldsymbol{\lambda}_{dyn}$ denotes the penalty coefficient vector for the dynamic security constraints, $\boldsymbol{\lambda}_{st}$ represents the penalty coefficient vector for the static security constraints, $\lambda_{opt}$ denotes the coefficient of generation cost, $\max(C)$ represents the maximum generation cost, and $\breve{\boldsymbol{V}}$, $\breve{\boldsymbol{P}}_G$, $\breve{\boldsymbol{Q}}_G$, and $\breve{\boldsymbol{P}}_L$ are the over-limit vectors of nodal voltage, active generation, reactive generation, and transmission power, respectively. Taking the nodal voltage as an example, $\breve{\boldsymbol{V}}$ is calculated as in (10):

$$\breve{\boldsymbol{V}} = \boldsymbol{max}\left(\boldsymbol{V} - \bar{\boldsymbol{V}}, \boldsymbol{0}\right) + \boldsymbol{max}\left(\underline{\boldsymbol{V}} - \boldsymbol{V}, \boldsymbol{0}\right) \tag{10}$$

where $\boldsymbol{max}()$ does element-wise comparisons and takes the maximum element at each position to form a result vector. Similarly, $\breve{\boldsymbol{P}}_G$, $\breve{\boldsymbol{Q}}_G$, and $\breve{\boldsymbol{P}}_L$ can be calculated.

On one hand, the dynamic simulator is implicitly incorporated into the MDP through the four-stage reward design. The features that describe the system dynamics are solely utilized for reward calculation and are not included in the state or observation. The environment conducts dynamic simulations and provides a reward, which indicates if the system meets the static and dynamic security constraints. The agent's objective is to maximize this reward. Consequently, the agent learns to generate actions that correspond to operation points satisfying both the static and dynamic security constraints.

On the other hand, to further illustrate the merits of the four-stage reward presented in (8), $\Delta T_S$ is compared with the maximum rotor angle difference at the end of the simulation $\Delta\delta_{\max}(T_E)$ and the Transient Stability Index $\alpha_{TSI}$ [27] in the IEEE 39-bus system. The variations of $\Delta\delta_{\max}(T_E)$, $\alpha_{TSI}$, and $\Delta T_S$ in relation to $P_{G\ 30}$ and $P_{G\ 38}$ are respectively displayed in Fig. 2. $\alpha_{TSI}$ is calculated as:

$$\left(\alpha_{TSI}; \gamma\right) = \frac{180 - \left(\Delta\delta_{\max}; \gamma\right)}{180 + \left(\Delta\delta_{\max}; \gamma\right)} \geq 0, \forall \gamma \in \boldsymbol{\Gamma} \tag{11}$$

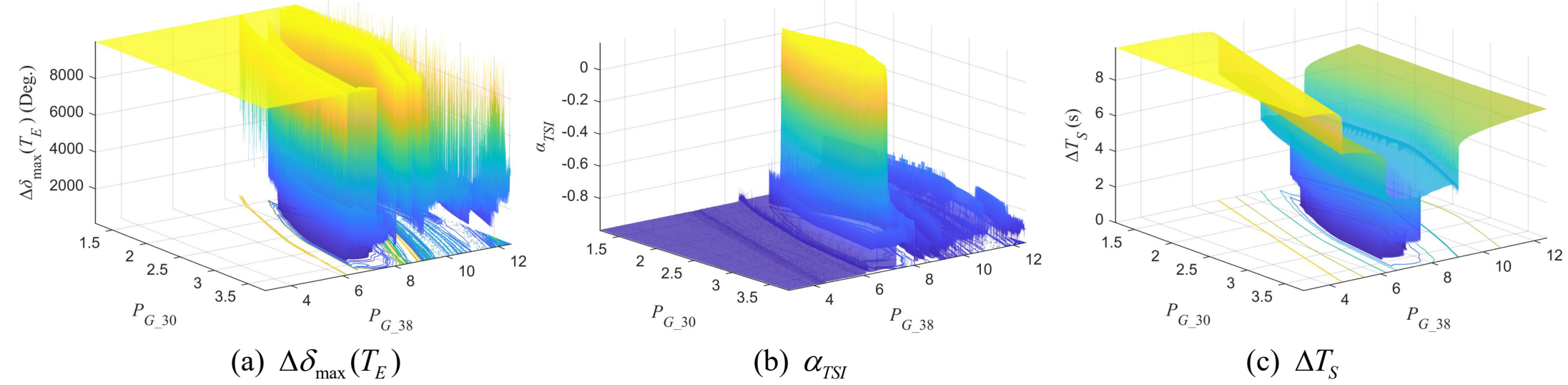

(a) $\Delta\delta_{\max}(T_E)$ (b) $\alpha_{TSI}$ (c) $\Delta T_S$

Fig. 2 Changes of $\Delta\delta_{\max}(T_E)$ , $\alpha_{TSI}$ , $\Delta T_S$ with respect to active power generations.

Compared with $\Delta\delta_{\max}(T_E)$ and $\alpha_{TSI}$ , $\Delta T_S$ mainly has three advantages. Firstly, $\Delta T_S$ offers valuable guiding information when agents are exploring the unstable domain by clearly quantifying the level of instability. Secondly, $\Delta T_S$ exhibits smoother changes and is strictly bounded within the range of $[0, T_E]$ , reducing the number of discontinuous points and the possibility of gradient anomalies. Finally, simulations can be terminated as soon as the power system is deemed unstable, which not only reduces simulation time but also significantly decreases the probability of non-convergent DAE solutions caused by severe instability.

Another advantage of adopting the instability duration-based reward is the flexibility to accommodate multiple dynamic constraints. Although only the rotor angle stability is considered in this paper, for a different stability constraint or multiple dynamic constraints, the proposed method can be easily applied without redesigning the reward by implementing the corresponding discrimination processes of these constraints in the simulation environment.

For the TSC-OPF problem in this paper, only the optimal solution is needed. Therefore, it is a single-step control, i.e., input $\boldsymbol{o} = [\boldsymbol{P}_D^T, \boldsymbol{Q}_D^T]^T$ and then output the optimum solution $\boldsymbol{a} = [\boldsymbol{V}_C^T, \boldsymbol{P}_C^T]^T$ . The discount factor $\gamma$ is set to 0. At this point, the objective of this MDP model is to maximize the expectation of the value of the initial state. This is equivalent to maximizing the expectation of reward $R_1$ , as shown in (12).

$$\max \quad v(\boldsymbol{s}_0) = \mathbb{E}(r_1 | \boldsymbol{s}_0), \boldsymbol{s}_0 \in [\boldsymbol{V}_C^T, \boldsymbol{P}_C^T, \boldsymbol{P}_D^T, \boldsymbol{Q}_D^T]^T \tag{12}$$

## IV. DDPG-CPEN FOR DRL-BASED TSC-OPF

### *A. Basics of the DDPG Algorithm*

The agent's policy $\pi$ is the basis of the agent choosing an action according to the observation. The DDPG algorithm is a widely used off-policy DRL algorithm. It adopts a deterministic policy, which is modeled as a mapping function shown in (13):

$$\boldsymbol{a} = \pi(\boldsymbol{a} | \boldsymbol{o}) = \boldsymbol{\mu}(\boldsymbol{o}) \tag{13}$$

The agent structure of the DDPG algorithm is displayed in Fig. 3. The actor, critic, target actor, and target critic neural networks are trained by minimizing the Temporal Difference (TD) error $[Q-(r+\gamma Q')]^2$ , where $Q$ is the action value. Since the discount factor $\gamma$ is zero, the TD error becomes $(Q-r)^2$ in this paper. The replay buffer contains transitions $(\boldsymbol{o}, \boldsymbol{a}, r, \boldsymbol{o}')$ , where $\boldsymbol{o}'$ is the subsequent observation after excuting action $\boldsymbol{a}$ . Details of the training procedures of the original DDPG algorithm can be found in [28].

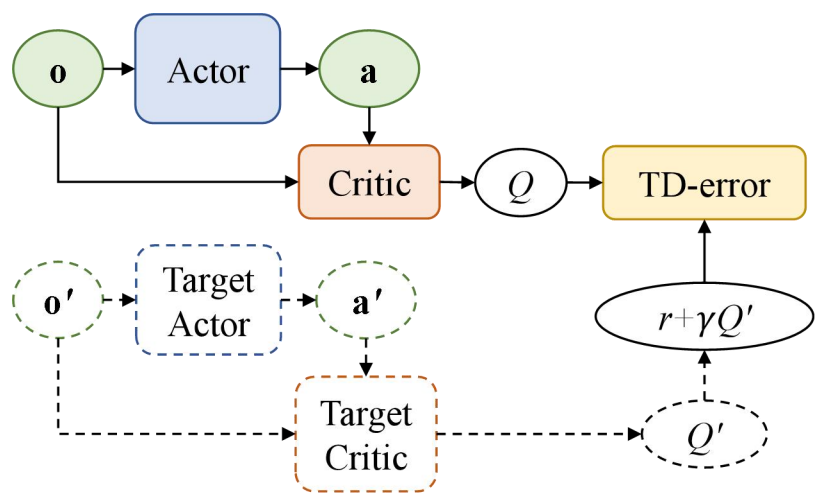

Fig. 3 Agent structure of the DDPG algorithm.

The main reason for choosing the DDPG algorithm is that off-policy DRL algorithms, such as DDPG, prioritize exploring the action space and training the agent with diverse samples, whereas on-policy DRL algorithms focus on utilizing and fine-tuning the existing policy. In the context of the TSC-OPF problem, the agent's generated action corresponds to a new power flow, which may not converge, especially during the early stages of training. If an on-policy DRL algorithm is employed, it is likely to require numerous training episodes to learn how to generate actions that result in convergent power flows, unless the initial policy can already generate such actions. In contrast, the off-policy algorithm can utilize samples from other policies for training, allowing us to incorporate a curriculum learning stage demonstrated in the subsequent part to mitigate this issue. Among the off-policy DRL algorithms, the DDPG algorithm is selected due to its adequate performance and simplicity.

### *B. Curriculum Learning*

In the initial phase of training, when the agent explores the action space with a large exploration rate, the selected actions can easily result in non-convergent power flows. This can severely hinder the efficiency and convergence of the agent's training. Therefore, in this paper, the idea of curriculum learning is introduced into the training procedures. Before learning how to generate actions that lead to appropriate operating points satisfying both static and dynamic security constraints, the agent initially focuses on learning how to generate actions that result in convergent power flows.

During the curriculum learning stage, two exploration schemes are employed alternately to fill the replay buffer. One scheme involves normal action selections with exploration noise as described in (14). The other scheme involves randomly selecting actions until a convergent power flow is obtained. A stepwise power flow sampling technique [29] is adopted to help the agent efficiently select actions corresponding to convergent power flows.

$$\boldsymbol{a} = \text{clip}\left(\boldsymbol{\mu}(\boldsymbol{o}) + \frac{\overline{\boldsymbol{a}} - \underline{\boldsymbol{a}}}{2} \odot \boldsymbol{\sigma}, \underline{\boldsymbol{a}}, \overline{\boldsymbol{a}}\right), \boldsymbol{\sigma} \sim \mathcal{N}(0, \epsilon) \tag{14}$$

where $\boldsymbol{o}$ denotes the observation vector, $\overline{\boldsymbol{a}}$ and $\underline{\boldsymbol{a}}$ are vectors representing the upper and lower bounds for the action respectively, $\epsilon$ denotes the exploration rate, $\boldsymbol{\sigma}$ represents the noise vector that follows a normal distribution with a mean of zero and a variance of $\epsilon$, and the clip function restricts the action with exploration noise within its upper and lower bounds.

The incorporation of the curriculum learning stage can enhance the ratio of samples associated with convergent power flows, thereby enhancing the efficiency of training the agent to generate actions that result in convergent power flows. On this basis, following the curriculum learning stage, the agent is then trained using regular procedures to gradually develop the ability to determine suitable operating points that minimize generation costs while satisfying static and dynamic security constraints.

### *C. Parallel Exploration*

To effectively cover the state space of power flow operations, a considerable number of samples is required, i.e., more training episodes are needed.

Therefore, a parallel exploration technique is adopted during the training process in this paper. Multiple TSC-OPF environments are created and action space explorations are performed concurrently. This parallel exploration approach enables the rapid filling of the replay buffer and enhances the efficiency of the agent in exploring the action space. Importantly, it is worth noting that the introduction of parallel exploration, unlike asynchronous DRL algorithms and multi-agent DRL algorithms, does not alter the original algorithm flow and is straightforward to implement. As a result, more samples can be obtained within the same number of training episodes.

Additionally, in this paper, the exploration rate $\epsilon$ in (14) is initially set to 1.0 and gradually decreases with training episodes. As the number of training episodes increases, $\epsilon$ linearly decreases until it reaches 0.1 and then remains unchanged, balancing exploration and exploitation.

### *D. Ensemble Decision-making*

After utilizing the improved MDP modeling and the above techniques, the agent training converges quickly and the performance of the agents obtained from multiple times of training is stable.

Therefore, an ensemble decision-making scheme is utilized that integrates the solutions of multiple well-trained agents. In this study, 5 agents are trained with the algorithm mentioned above. After the training is completed, when making decisions in the application stage, the strategy with the maximum reward is selected for execution. The pseudocode of the proposed DDPG-CPEn algorithm is illustrated below.

**The DDPG-CPEn Algorithm**

**Agent Training**

**Input**: the actor network $\boldsymbol{\mu}(;\boldsymbol{\xi})$ and its learning rate $\alpha_{\xi}$, the critic network $Q(;\boldsymbol{\theta})$ and its learning rate $\alpha_{\theta}$, the total number of training episodes $T_{episode}$, the number of episodes in the curriculum learning stage $T_{clm}$, the number of threads $N_P$, the replay buffer $\mathcal{B}$, and the mini-batch size $N_{batch}$ of agent training.

Randomly initialize the parameters $\boldsymbol{\theta}$ and $\boldsymbol{\xi}$ of actor and critic.
Initialize the replay buffer $\mathcal{B}$.
Initialize the exploration rate: $\epsilon \leftarrow 1.0$.
Generate $N_P$ threads and create $N_P$ TSC-OPF environments.
**for** $\tau = 1$ **to** $T_{episode}$ **do**
  Concurrently, reset the environments and sample a state $\boldsymbol{s}$ in the custom state space $[\boldsymbol{V}_C^T, \boldsymbol{P}_C^T, \boldsymbol{P}_D^T, \boldsymbol{Q}_D^T]^T$ that leads to convergent power flow.
  The agent gets observations $\boldsymbol{o}$ from the $N_P$ environments.
  **if** $\tau < T_{clm}$ **and** $\tau \bmod 2 == 1$ **do**
    The agent randomly selects $N_P$ actions corresponding to convergent power flows using the stepwise power flow sampling technique [29] and sends them to the environments.
  **else do**
    The agent selects $N_P$ actions according to (14) and distributes them to the environments.
**end if**
Concurrently, each environment executes the corresponding action, calculates the reward $r$, and transitions to a new state $\boldsymbol{s}'$.
The agent gets $N_P$ transitions $(\boldsymbol{o}, a, r, \boldsymbol{o}')$ from environments and stores them in the replay buffer $\mathcal{B}$.
Randomly sample $N_{batch}$ transitions from $\mathcal{B}$.
Update the critic network by:

$$\boldsymbol{\theta} \leftarrow \boldsymbol{\theta} - \alpha_{\theta} \frac{1}{N_{batch}} \sum_{k=1}^{N_{batch}} \nabla_{\theta} \left[Q(\boldsymbol{o}, \boldsymbol{a}; \boldsymbol{\theta}) - r\right]^2$$

Update the actor network by:

$$\boldsymbol{\xi} \leftarrow \boldsymbol{\xi} + \alpha_{\xi} \frac{1}{N_{batch}} \sum_{k=1}^{N_{batch}} \nabla_{\xi} \left\{Q\left[\boldsymbol{o}, \boldsymbol{\mu}(\boldsymbol{o}; \boldsymbol{\xi}); \boldsymbol{\theta}\right]\right\}$$

Update the exploration rate: $\leftarrow \max(\ -1/T_{epoch}, 0.1)$.
**end for**

**Agent Application**

Input: five well-trained agents and current operating point.

Input observation $\boldsymbol{o}$ to the actor networks of five agents and get five actions.
Calculate the rewards of five actions in the TSC-OPF environment.
Choose the action with the maximum reward as the final solution.

## V. CASE STUDY

The IEEE 39-bus system and a practical 710-bus regional power grid are utilized for numerical experiments. In both cases, dynamic components are modeled with detailed models. The adopted generator model is the sixth-order model with different kinds of excitation controllers, governors, and power system stabilizers. A composite load model of the induction motor and constant impedance is used. The total simulation time $T_E$ is 10 seconds and the time step is 0.01 seconds. Simulations will terminate whenever the maximum rotor angle difference exceeds 500 degrees. When sampling power flows, the bus voltage and active generation of each generator are arbitrarily selected within their respective upper and lower limits. Moreover, the active and reactive loads are selected arbitrarily within the range of 0.7 to 1.2 times the given load level.

Using the open-source power system time-domain simulator Py_PSOPS [29], the environments for TSC-OPF are established. The DDPG-CPEn algorithm is programmed with Python. Parallel exploration is realized using a Python library called ray [30]. The testing platform is a high-performance server installed with the Linux operating system. The server is equipped with one Intel i7-10700KF 3.80 GHz octa-core CPU processor, supporting 16 threads after enabling hyper-threading technology, one Nvidia RTX 3090 GPU processor, and 128GB DDR4-3200MHz RAM. The complete implementation of the proposed method has been made publicly available on GitHub.

Table I. Designs of agents.

| Test case | Method | Actor (MLP) | Critic (MLP) |
|---|---|---|---|
| IEEE-39 | Base | **97**-256×3-19 | **116**-256×3-1 |
| | Proposed | **38**-256×3-19 | **57**-256×3-1 |
| 710-Bus | Base | **1146**-512×3-117 | **1263**-512×3-1 |
| | Proposed | **318**-512×3-117 | **435**-512×3-1 |

Table II. Settings of hyperparameters.

| Test case | Methods | $\lambda_{opt}$ | $\alpha_{\xi}$ | $\alpha_{\theta}$ | $N_{batch}$ | $T_{epoch}$ | $N_P$ | $T_{clm}$ |
|---|---|---|---|---|---|---|---|---|
| IEEE-39 | Base | 2000 | 1.0e-4 | 0.001 | 256 | 20,000 | **1** | **0** |
| | Proposed | | | | | | **10** | **2,000** |
| 710-Bus | Base | 2000 | 1.0e-5 | 0.001 | 1024 | 50,000 | **1** | **0** |
| | Proposed | | | | | | **10** | **5,000** |

Multi-Layer Perceptron (MLP) is used to build actor and critic neural networks of agents. The designs of agents and the settings of hyperparameters are demonstrated in Table I and Table II, respectively.

In these tables, the term "**Base**" refers to the original DDPG algorithm with MDP modeling, where the whole power flow state is used as the observation and the reward design is based on $\Delta\delta_{\max}(T_E)$. The $\Delta\delta_{\max}(T_E)$-based reward design differs from the proposed $\Delta \boldsymbol{T}_S$-based reward design shown in (8) only when the state violates dynamic constraints, as depicted in (15). On the other hand, the term "Proposed" refers to the proposed fast-converging DRL method that utilizes the DDPG-CPEn algorithm with improved MDP modeling.

$$R = \max\left(-500 - \lambda_{dyn}^{T} \min\left(\Delta\delta_{\max}(T_E) - 180, 500\right), -999\right) \quad (15)$$

In Table I, due to the observation space reduction, it can be observed that the neural networks of the proposed method are significantly smaller than those of the Base method. On the other hand, the determination of hyperparameters, as shown in Table II, involves the following procedures. Firstly, the coefficient of generation cost $\lambda_{opt}$ is selected so that the positive reward value is around 1,000 in most cases. Secondly, the values of $N_{batch}$ and $T_{episode}$ are set as 256 and 50,000 respectively, followed by a grid search on the learning rates $\alpha_{\xi}$ and $\alpha_{\theta}$. This search helps identify the learning rates that yield the best-performing agent. The learning rate for the actor network $\alpha_{\xi}$ is chosen from the values 1.0e-3, 1.0e-4, and 1.0e-5, while the learning rate for the critic network $\alpha_{\theta}$ is chosen from 1.0e-2, 1.0e-3, and 1.0e-4. Thirdly, the mini-batch size $N_{batch}$ is increased if the agent's performance exhibits excessive fluctuations. Finally, the total number of training episodes $T_{episode}$ is reduced while ensuring that the performance of well-trained agents is maintained. Regarding $N_P$ and $T_{clm}$, the proposed method selects the number of threads $N_P$ to be 10 and the number of episodes in the curriculum learning stage $T_{clm}$ to be one-tenth of $T_{episode}$. The volume limit of the replay buffer $\mathcal{B}$ is set to 1.0e6, ensuring that all historical state transition experiences are stored and utilized for agent training.

It is worth noting that, in order to assess the convergence of agent training and the stability of the obtained agents' performance, five agents are trained using random seeds of 1024, 2048, 3072, 4096, and 5120 for each DRL method. The convergence speed and model performance of the five agents are compared. These agents are used for the ensemble decision-making of the proposed method.

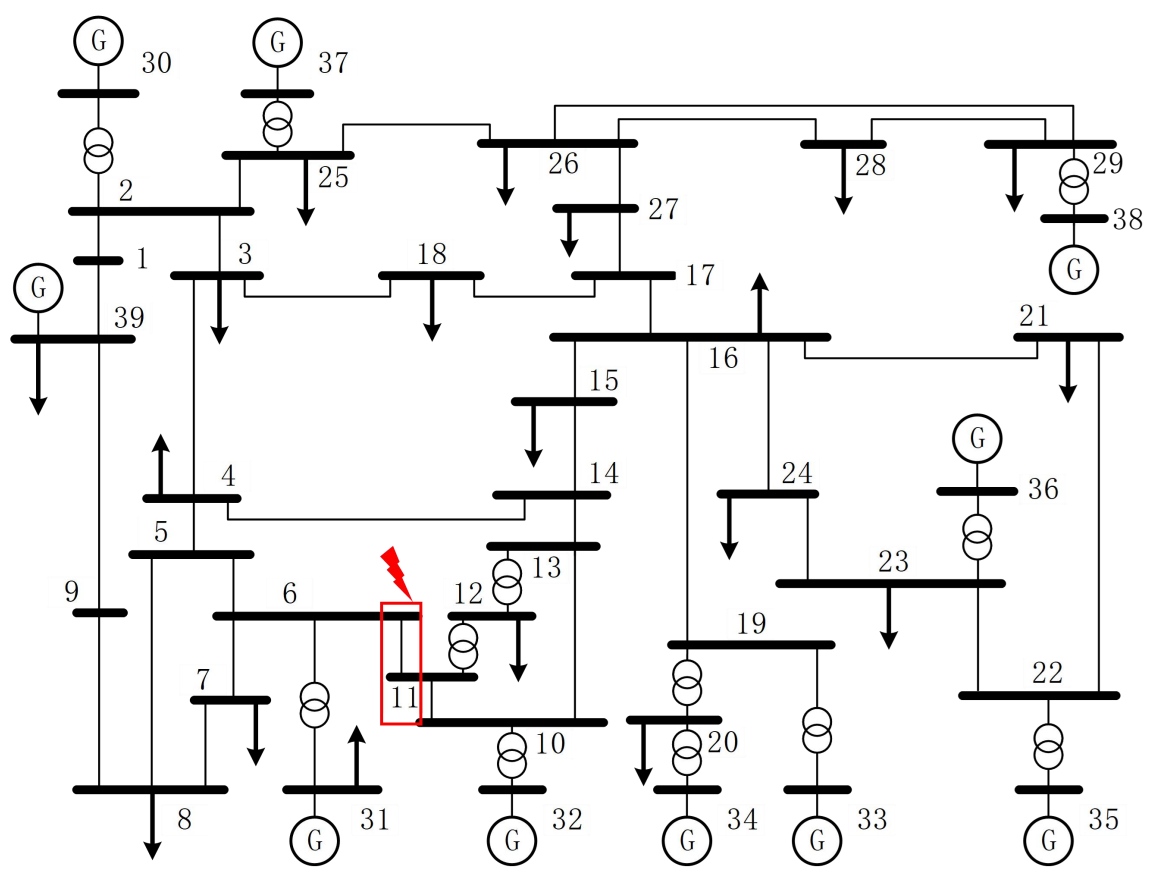

Fig. 4 The IEEE 39-bus system and the anticipated contingencies.

## *A. IEEE-39 System*

In the IEEE 39-bus system, there are two anticipated contingencies including the three-phase short circuit fault at the first end and the last end of the line between bus 6 to bus 12, which are cleared after 0.1 seconds. The topology of the IEEE 39-bus system and the fault location are illustrated in Fig. 4.

### *1) Comparisons with the Base Method*

As shown in Table I, the observation space for the Base method comprises the voltage amplitudes of 39 buses, the active and reactive power outputs of 10 generators, and the active and reactive power consumptions of 19 loads. The action space includes the nodal voltages of 10 generator buses and the active generations of 9 generators, as there is one slack machine. As a result, the observation space has 97 dimensions, while the action space has 19 dimensions.

In terms of the agent, the actor network consists of an input layer with 97 dimensions, three hidden layers of 256 dimensions each, and an output layer with 19 dimensions. On the other hand, the Critic network is made up of an input layer with 116 dimensions, three hidden layers of 256 dimensions each, and an output layer with 1 dimension.

As for the proposed method, the observation space is reduced to 38 dimensions, which only contains the active and reactive power consumptions of 19 loads. The number of episodes of the curriculum learning stage $T_{clm}$ is set to 2,000. The number of threads $N_P$ is set to 10, i.e., 10 independent environments, and 10 threads are created for parallel exploration.

During the training process, agents are evaluated every 100 episodes by the average reward of 100 random scenarios. As previously mentioned, for each method, five agents are trained with different random seeds, which means that five evaluation curves can be drawn. By overlapping the five evaluation curves, the training processes of the Base method and the proposed method are compared in Fig. 5. The term "Avg" represents the average of five evaluation curves, while "MinMax" represents the range covered by the curves. The comparative results

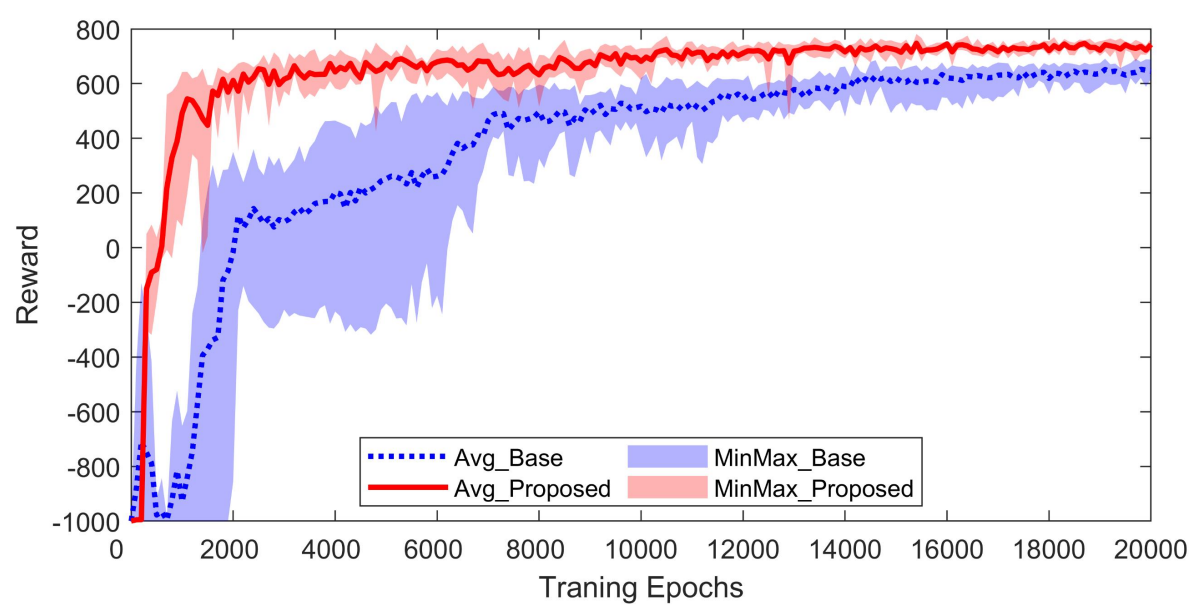

Fig. 5 Training processes of the Base method and the proposed method in the IEEE 39-bus system.

Table III. Agent performance within 10,000 new scenarios in the IEEE 39-bus system.

| Method | Avg_r | S% | F_S% | F_D% | F_NC% |
|---|---|---|---|---|---|
| Base | 642.05 | 97.91% | 2.03% | 0.04% | 0.02% |
| Proposed | **740.23** | **99.80%** | **0.01%** | **0.19%** | **0.00%** |

"Avg_r" denotes the average reward, "S%" represents the rate of solutions that maintain static and dynamic security, i.e., the success rate, "F_S%" denotes the rate of static security constraint violations, "F_D%" represents the rate of dynamic security constraint violations, and "F_NC%" denotes the rate of non-convergent power flows.

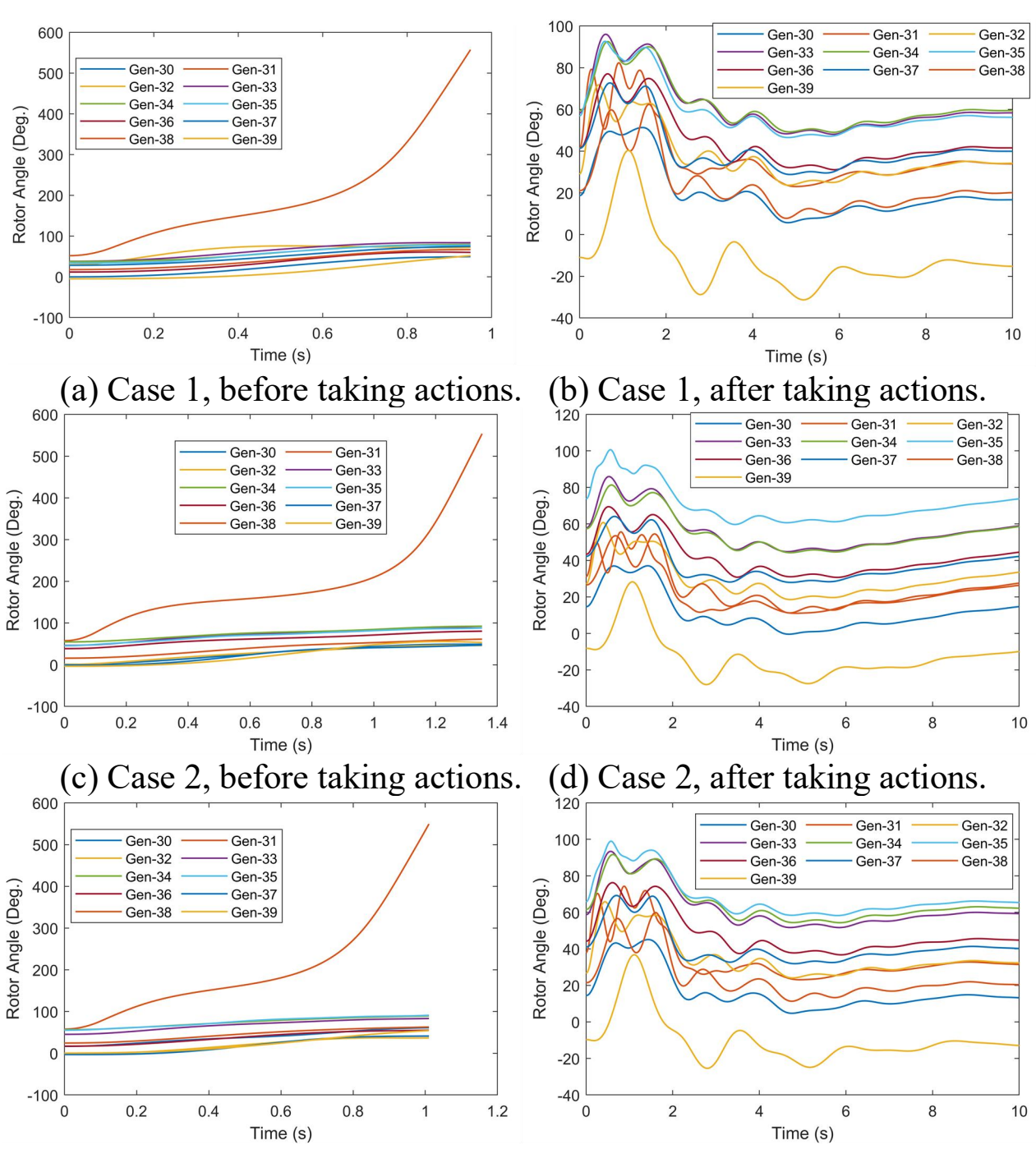

(a) Case 1, before taking actions. (b) Case 1, after taking actions.

(c) Case 2, before taking actions. (d) Case 2, after taking actions.

(e) Case 3, before taking actions. (f) Case 3, after taking actions.

Fig. 6 Rotor angle curves before and after taking the actions generated by the agents in the IEEE 39-bus system.

illustrate that the proposed algorithm not only exhibits faster and more stable convergence but also achieves higher rewards compared to the Base algorithm.

After training, 10,000 new scenarios that violate dynamic security constraints are generated with a random seed of 42. A successful action is an action that corresponds to an operation point within the feasible region where all the static security

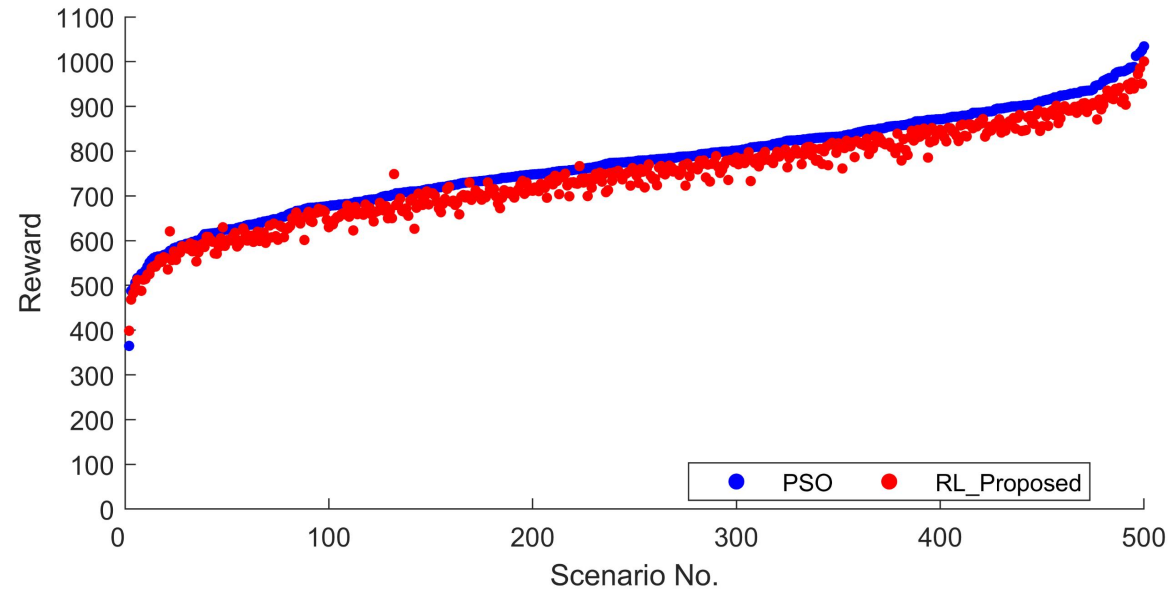

Fig. 7 Rewards of the proposed method and the PSO algorithm in the IEEE 39-bus system.

Table IV. Agent performance and time consumption of the proposed method and the PSO algorithm in the IEEE 39-bus system.

| Method | T_T (hour) | N_S | T_D (hour) | S% | R% |
|---|---|---|---|---|---|
| PSO | N/A | 500 | 82.68 | 100.00% | 100.00% |
| Proposed | **4.73** | | **0.017** | **99.80%** | **96.12%** |

"T_T" represents the time required for agent training, "N_S" denotes the number of scenarios used for testing, "T_D%" represents the time spent on decision-making and solution evaluation, and "R%" denotes the average ratio of rewards obtained by the proposed method compared to those of the PSO algorithm.

constraints and dynamic constraints are not violated. The comparative results obtained from these tests are presented in Table III. A higher average reward, a higher success rate S%, and lower rates of static security constraint violations F_S%, dynamic security constraint violations F_D%, and non-convergent power flows F_NC% demonstrate the superior performance of the proposed method. The rotor angle curves of three test scenarios are demonstrated in Fig. 6. After taking the actions generated by the agents, the power system can maintain rotor angle stability.

*2) Comparisons with the PSO Algorithm*

The performance of the proposed method is compared with the PSO algorithm, which is implemented using the scikit-opt package in Python. The PSO algorithm employs 200 particles and runs for 150 iterations. Consequently, for each TSC-OPF scenario, 30,000 power flow solutions and 60,000 stability simulations are performed.

Both the PSO algorithm and the decision-making procedures of the proposed method are executed using a single CPU core, and the time costs are recorded. Due to the substantial time requirement of PSO, only 500 random scenarios that violate dynamic security constraints are utilized for the comparison. The results are displayed in Fig. 7 and Table IV.

In terms of time consumption, the PSO algorithm requires 82.68 hours to search for solutions for the 500 scenarios, equivalent to about 9.92 minutes per scenario. In contrast, it takes 4.73 hours to train five agents using the proposed method. Determining and evaluating solutions for the 500 scenarios takes 0.017 hours, which is approximately 0.12 seconds per scenario. The execution time per scenario is only about 0.02 percent of that using the PSO algorithm. Importantly, the time

Table V. Performance comparisons of 10,000 new scenarios in the IEEE 39-bus system.

| Method | Avg_r | S% | F_S% | F_D% | F_NC% |
|---|---|---|---|---|---|
| $\Delta\delta_{\max}$ (Base) | 642.05 | 97.91% | 2.03% | 0.04% | 0.02% |
| $\alpha_{TSI}$ | 665.57 | 97.82% | 2.14% | 0.04% | 0.00% |
| $\Delta T_S$ (AllState) | 690.07 | 98.19% | 1.65% | 0.16% | 0.00% |
| OnlyLoad (Origin) | 700.02 | 98.30% | 1.11% | 0.59% | 0.00% |
| Curriculum | 689.32 | 98.62% | 0.20% | 1.18% | 0.00% |
| Parallel | 699.19 | 98.95% | 0.62% | 0.43% | 0.00% |
| DDPG_CP | 724.84 | 99.26% | 0.08% | 0.66% | 0.00% |
| Proposed | **740.23** | **99.80%** | **0.01%** | **0.19%** | **0.00%** |

"$\Delta\delta_{\max}$ (Base)" represents the Base method, "$\alpha_{TSI}$" denotes the modified Base method with a $\alpha_{TSI}$-based reward design shown in (16), "$\Delta T_S$ (AllState)" represents the modified Base method with the $\Delta T_S$-based reward design shown in (8), "OnlyLoad (Origin)" denotes the method that introduces the improved MDP modeling into the "$\Delta T_S$ (AllState)" method, "Curriculum" and "Parallel" represents the methods that introduce curriculum learning and parallel exploration into the "OnlyLoad (Origin)" method, respectively, and "DDPG_CP" denotes the proposed method without ensemble decision-making.

cost of the PSO algorithm far surpasses the total time consumed by agent training and decision-making. This gap will continue to widen as the number of test scenarios increases.

In terms of model performance, the PSO algorithm demonstrates a 100.00 percent success rate in decision-making and commonly achieves better rewards. On the other hand, the proposed method achieves a success rate of 99.80 percent in decision-making, with only one failure observed in 500 scenarios. Considering the remarkably fast decision-making speed, it is possible to switch to alternative methods once the agent's solution has been verified as unsuccessful. Additionally, the average reward attained by the proposed method is 96.12 percent of that achieved by the PSO algorithm. In Figure 6, several scenarios are illustrated where the proposed method outperforms the PSO algorithm.

Overall, the proposed method considers both accuracy and efficiency when solving TSC-OPF. While achieving fast decision-making speed, it only incurs a slight decrease in performance compared to the PSO algorithm.

*3) Ablation Experiments*

Ablation experiments are conducted to test the effects of the proposed improvements to the MDP modeling and the DDPG algorithm. The results are presented in Table V. The $\alpha_{TSI}$-based reward design deviates from the proposed $\Delta T_S$-based reward design, as described in (8), only when the state violates dynamic constraints, as illustrated in (16).

$$R = \max\left(-500 + \lambda_{dyn}^{T}\alpha_{TSI}, -999\right) \quad (16)$$

Firstly, three reward designs are compared and the corresponding comparative results are shown in Fig. 8 and the first three lines of Table V. When compared to the $\Delta\delta_{\max}(T_E)$-based and $\alpha_{TSI}$-based reward designs, the $\Delta T_S$-based reward design improves agent training by facilitating faster convergence and results in better agent performance.

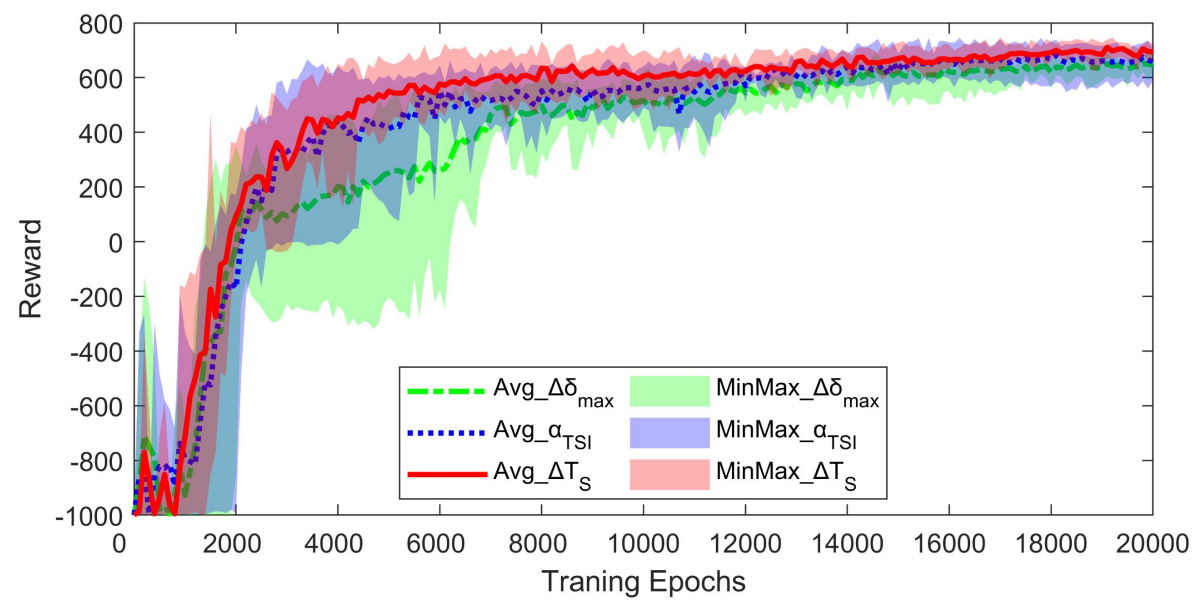

Fig. 8 Training processes of methods with different reward designs.

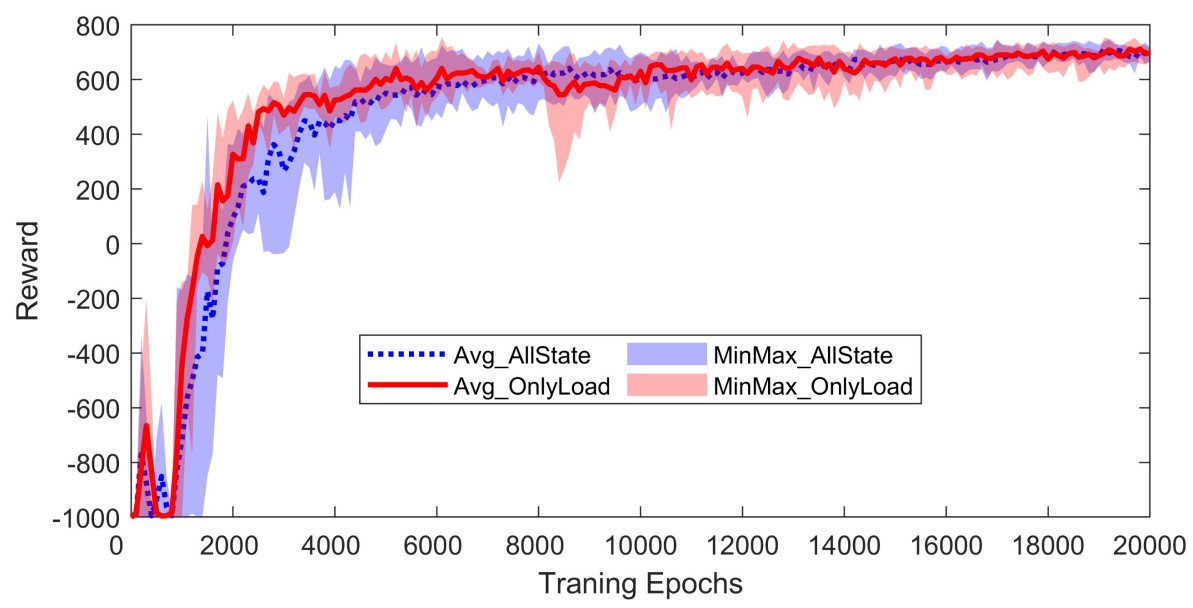

Fig. 9 Training processes of methods with different observation designs.

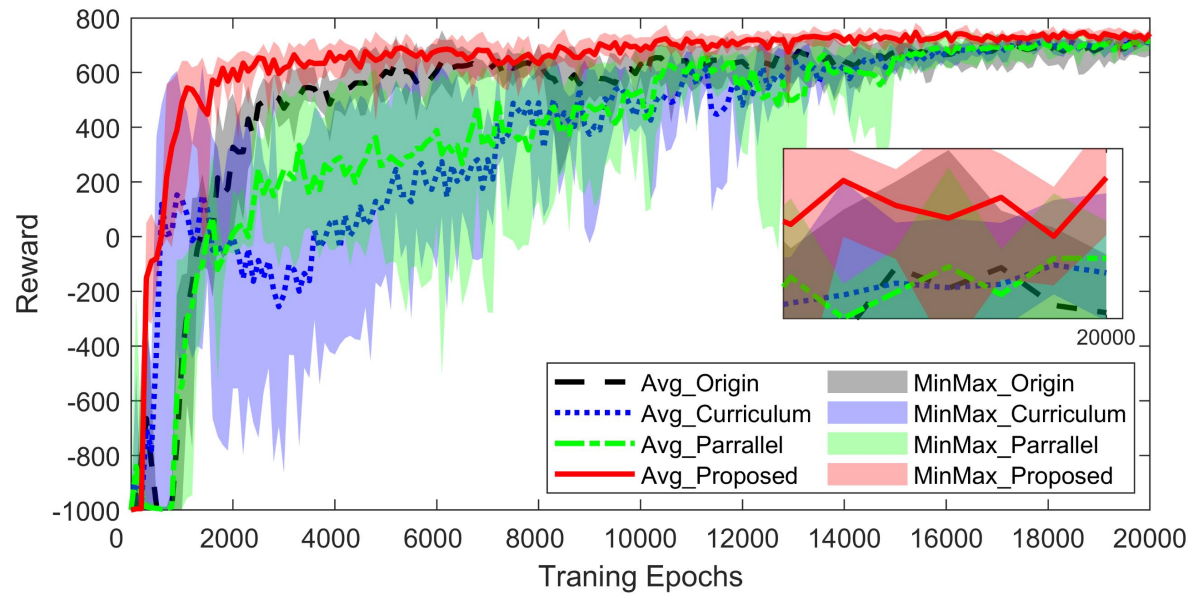

Fig. 10 Training processes of methods with different training algorithms.

Secondly, two observation designs are compared. The results are presented in Fig. 9 and the third and fourth lines of Table V. The results demonstrate that reducing the observation space leads to faster convergence during the early training stage, without compromising the final performance of the agent.

Finally, different training algorithms are compared, and the results are displayed in Fig. 10 and lines four to eight of Table V.

By adding a curriculum learning phase, the efficiency of the agent learning how to generate actions that lead to convergent power flows during the early training stage is significantly improved. However, completing the curriculum learning phase results in a sudden change in the training procedures, which also has an impact on the stability of agent training to some extent.

The parallel exploration enhances the efficiency of the agent in exploring the action space. However, it also influences the early training stage of the agent. In the absence of curriculum learning, parallel exploration during the early training stage results in a higher number of non-convergent power flow samples being stored in the replay buffer. This, in turn, impacts the agent's ability to learn how to generate actions that lead to convergent power flows, reducing its learning efficiency.

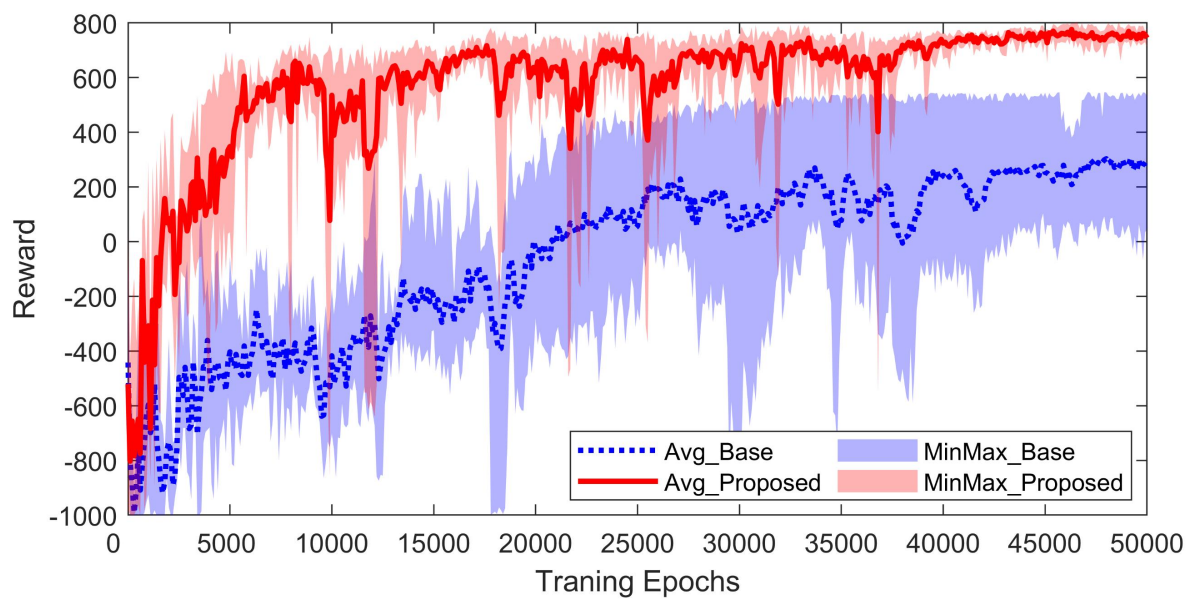

Fig. 11 Training processes of the Base method and the proposed method in the practical 710-bus power grid.

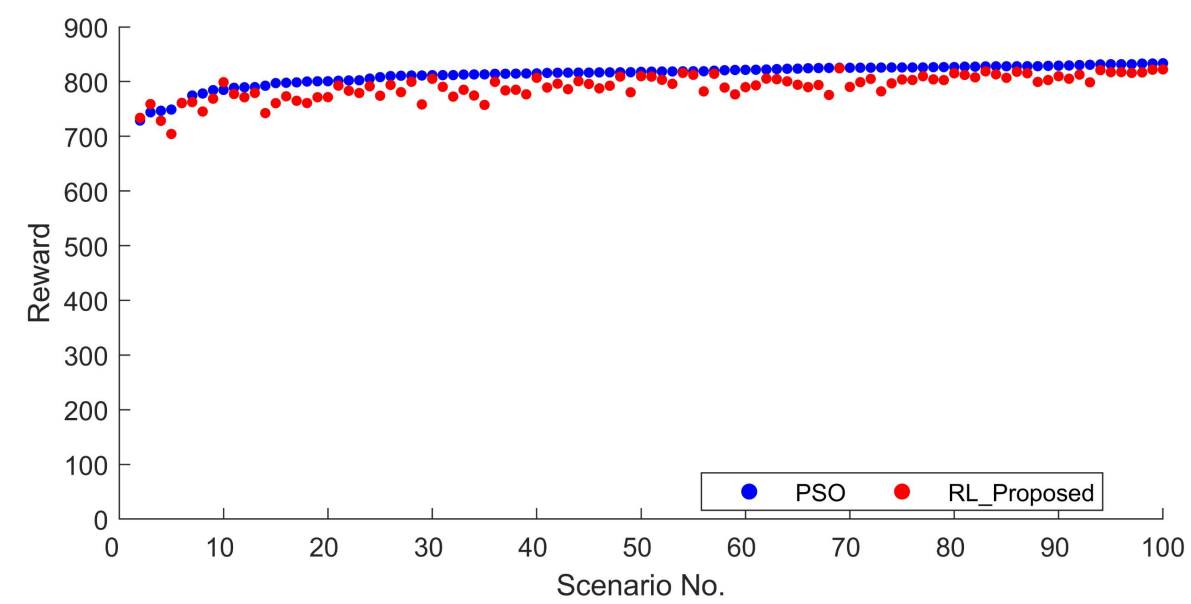

Fig. 12 Rewards of the proposed method and the PSO algorithm in the practical 710-bus power grid.

Table VI. Agent performance within 10,000 new scenarios in the practical 710-bus power grid.

| Method | Avg_r | S% | F_S% | F_D% | F_NC% |
| --- | --- | --- | --- | --- | --- |
| Base | 198.53 | 82.26% | 0.27% | 16.77% | 0.70% |
| Proposed | **790.43** | **100.00%** | **0.00%** | **0.00%** | **0.00%** |

Generally, a large $T_{episode}$ results in an adequate number of state transition samples stored in the replay buffer, leading to a well-trained agent with satisfactory performance. However, the TSC-OPF problem poses a challenge due to the sparse reward issue, requiring more exploration episodes during agent training to obtain sufficient samples. By combining the curriculum learning stage and parallel exploration, the replay buffer can be efficiently filled with a large number of samples that comprehensively cover both the state space and the action space. Notably, these samples also encompass a high proportion of samples that correspond to convergent power flows and lie in the feasible region of the TSC-OPF problem. Consequently, the DDPG-CP algorithm and the proposed method share the same training process in Fig. 10, which converges rapidly and stably. After training, the proposed method further enhances the agent by leveraging the ensemble decision-making scheme.

### *B. 710-Bus Practical Regional Power Grid*

In the practical 710-bus power grid, there are two anticipated contingencies including the three-phase short circuit fault at the first end and the last end of the line between bus 83 to bus 153, which are cleared after 0.1 seconds. The agent designs and hyperparameter settings are displayed in Table I and Table II.

#### *1) Comparisons with the Base Method*

As for the Base method, the observation space comprises the voltage amplitudes of 710 buses, the active and reactive power outputs of 59 generators, and the active and reactive power consumptions of 159 loads. The action space includes the nodal voltages of 59 generator buses and the active generations of 58 generators, as there is one slack machine. As a result, the observation space has 1146 dimensions, while the action space has 117 dimensions. As for the proposed method, the observation space is reduced to 318 dimensions consisting of the active and reactive power consumptions of 159 loads.

Table VII. Agent performance and time consumption of the proposed method and the PSO algorithm in the IEEE 39-bus system.

| Method | T_T (hour) | N_S | T_D (hour) | S% | R% |
| --- | --- | --- | --- | --- | --- |
| PSO | N/A | 100 | 163.79 | 100.00% | 100.00% |
| Proposed | **78.39** | | **0.033** | **100.00%** | **97.32%** |

Comparative results are depicted in Fig. 11 and Table VI. As the power grid expands, the observation space, the actor space, and the scale of neural networks increase significantly, amplifying the sparse reward problem. The Base method requires a substantial number of exploration episodes, yet still struggles to acquire sufficient positive rewards. Consequently, it exhibits slow convergence, unstable training processes, and unsatisfactory agent performance. Conversely, the proposed method consistently yields five superior agents after five times of training. The training processes remain rapid and stable.

It is worth noting that in comparison to the IEEE 39-bus system, the 710-bus power grid exhibits even sparser positive rewards. With an adequate number of positive reward samples, the training effectiveness of the agents can be enhanced. Among the newly sampled 10,000 dynamic insecure scenarios, the proposed method achieves a solution effectiveness of 100.00 percent.

It is important to mention that in the case of the DRL method, the performance of agents tends to be better in smaller systems compared to larger systems. However, when considering the success rate, the proposed DRL method may achieve a higher success rate in larger systems. In the proposed DRL method, each well-trained agent will cover a specific range within the feasible region. During the application stage, it is theoretically possible to cover the entire feasible region by having a sufficient number of agents. This phenomenon occurs because even though the ratio of the feasible region to the overall operation region of a large power system is very small, resulting in a more severe sparse reward problem and more challenging training process, the smaller feasible region may also be relatively easier to cover by the combined efforts of multiple well-trained agents. The premise is to have a sufficient number of well-performing agents, which is facilitated by the proposed DRL method. This is demonstrated in the 710-bus system, where the policy functions of the five agents have successfully covered the entire feasible domain.

*2) Comparisons with the PSO Algorithm*

Similarly, the proposed method is compared with the PSO algorithm in the 710-bus power grid. The PSO algorithm still employs 200 particles and runs for 150 iterations.

Due to the time-consuming nature of PSO-based TSC-OPF solutions in the 710-node system, the number of test scenarios is reduced to 100. Comparative results can be found in Figure 11 and Table VII. Just like in the IEEE 39-bus system, the PSO algorithm requires more computational costs but achieves better solutions. The solutions are 100.00 percent effective, and it takes the PSO algorithm 163.79 hours to search for solutions for the 100 scenarios, which is about 1.64 hours per scenario.

On the other hand, the proposed method also achieves 100.00 percent solution effectiveness. Determining and evaluating solutions for the 100 scenarios only takes 0.033 hours, i.e., approximately 1.18 seconds per scenario, which is only about 0.02 percent of that using the PSO algorithm. Meanwhile, the average reward is 97.32% of that of the PSO algorithm. The total time consumption of training five agents and performing decision-making is also shorter than that of the PSO algorithm. These results further validate the effectiveness and efficiency of the proposed method.

## VI. Conclusions

In this paper, a fast-converging DRL-based PSOD-TSC solution method is proposed. The aim is to address the sparse reward problem and enhance the exploration efficiency of DRL agents. The proposed method includes an improved MDP modeling of TSC-OPF and an enhanced DDPG-CPEn algorithm, which contributes to stable and rapid agent training. The effectiveness and efficiency of the proposed method are validated through performance comparisons and ablation tests conducted on the IEEE 39-bus system and a practical 710-bus regional power grid. Remarkably, this method represents the first application of a DRL-based PSOD-TSC solution on such a large-scale power grid. The source code of the proposed method is available on GitHub. It is worth noticing that the proposed improvements on the MDP modeling and the training algorithm can be effective on other off-policy algorithms as well. It is important to note that the proposed enhancements to the MDP modeling and the training algorithm are also effective for other off-policy algorithms.

There are three directions for future work. Firstly, the design and training of agents that can adapt to topological changes should be considered. Currently, agents are trained assuming that the power system's topology will remain unchanged. As a result, the agent can adapt to changes in operation points but may not necessarily adapt to changes in topology. Studies on graph neural networks are ongoing [31]. Secondly, it is important to consider the long-term consequences of actions in the MDP modeling. The original TSC-OPF modeling in this paper does not account for temporal relationships, such as ramping constraints. Although the sequential TSC-OPF problem is challenging, it holds greater practical value. Lastly, the proposed method should be extended to distributed multi-agent DRL algorithms to further enhance exploration efficiency and enable the application of DRL-based solution methods to a wider range of power system scales.

**Tannan Xiao** received the B.Eng. and Ph.D. degrees in electrical engineering from Zhejiang University, Hangzhou, China, in 2013 and 2019, consecutively. From 2020 to 2023, he served as a Postdoctoral Fellow in the Department of Electrical Engineering at Tsinghua University, Beijing, China. Currently, he holds the position of Assistant Researcher at the State Key Laboratory of Power System Operation and Control, Tsinghua University, Beijing, China. His research interests include power system stability analysis and control, high-performance computing, and artificial intelligence.

**Ying Chen** received the B.S. and Ph.D. degrees in electrical engineering from Tsinghua University, Beijing, China, in 2001 and 2006, respectively, where he is currently a Professor with the Department of Electrical Engineering. His research interests include parallel and distributed computing, electromagnetic transient simulation, cyber-physical system modeling, and cyber security of smart grids.

**Han Diao** received his B.Eng. degree in Electrical Engineering from Tsinghua University in 2022. He is now pursuing the M.Eng. degree in the same department, at the Security Control and Efficient Utilization of Modern Power and Energy Systems Laboratory. His research focuses on applying artificial intelligence to power system stability analysis and control.

**Shaowei Huang** received the B.S. and Ph.D. degrees from Tsinghua University, Beijing, China, in 2006 and 2011, respectively. From 2011 to 2013, he was a Postdoctoral Fellow with the Department of Electrical Engineering, Tsinghua University, where he is currently an Associate Professor. His research interests include power systems modeling and simulation, power system parallel and distributed computing, complex systems and their application in power systems, and artificial intelligence.

**Chen Shen** received the B.E. and Ph.D. degrees in electrical engineering from Tsinghua University, Beijing, China, in 1993 and 1998, respectively. From 1998 to 2001, he was a Postdoc Research Fellow with the Department of Electrical Engineering and Computer Science, University of MissouriRolla, Rolla, MO, USA. From2001 to 2002, he was a Senior Application Developer with ISO New England, Inc., MA, USA. Since 2009, he has been a Professor with the Department of Electrical Engineering, Tsinghua University. He is currently the Director with Energy Digitization Research Center, Sichuan Energy Internet Research Institute, Tsinghua University. He is the author/coauthor of more than 200 technical papers and one book, and holds 50 issued patents. His research interests include power system analysis and control, renewable energy generation, and smart grids.